\documentclass{JAC2003}
\newcommand{\He}{\ensuremath{^6{\mathrm{He}\,}}}
\newcommand{\Ne}{\ensuremath{^{18}{\mathrm{Ne}\,}}}

\usepackage{graphicx}

\begin{document}
\title{Perspectives for a neutrino program based on the upgrades of
the CERN accelerator complex}

\author{A. Donini, I.F.T. and Dep. F\'{\i}sica Te\'{o}rica, U.A.M., Madrid, Spain\\
E. Fernandez, I.F.T. and Dep. F\'{\i}sica Te\'{o}rica, U.A.M., E-28049, Madrid,
Spain\\ P. Migliozzi\thanks{migliozzi@na.infn.it}, Istituto Nazionale
di Fisica Nucleare, Sezione di Napoli, Italy\\ S. Rigolin, I.F.T. and
Dep. F\'{\i}sica Te\'{o}rica, U.A.M., Madrid, Spain\\ L. Scotto Lavina, Dip. di
Fisica, Universit\`{a} "Federico II" and INFN, Napoli, Italy\\ T.
Tabarelli de Fatis, Universita di Milano Bicocca and INFN, Milano,
Italy\\ F. Terranova, Istituto Nazionale di Fisica
  Nucleare, Laboratori Nazionali di Frascati}

\maketitle

\begin{abstract}
In this paper, we discuss the possibilities offered to neutrino
physics by the upgrades of the CERN accelerator complex. Emphasis is
on the physics reach of a medium $\gamma$ (350-580) $\beta$-beam that
fully exploits the improvements in the CERN accelerator complex for
the luminosity/energy upgrade of the LHC. We show that,
this design not only profits of the ongoing efforts for the upgrades
of the LHC, but also leverage out the existing infrastructures of the
LNGS underground laboratory.  Furthermore, given the involved high
neutrino energies, above 1~GeV, a non-magnetized iron detector could
efficiently exploit the neutrino beam.

We show that the performance of this complex for what concerns the
discovery of the CP violation in the leptonic sector, in case
$\theta_{13}$ is discovered by Phase I experiments,  is comparable
with the current baseline design based on a gigantic water Cherenkov
at Frejus. Furthermore, this complex has also some sensitivity to the
neutrino mass hierarchy.
\end{abstract}

\section{Introduction}

The hypothesis of neutrino oscillations~\cite{Pontecorvo:1957yb} is
strongly supported by atmospheric~\cite{atmo}, solar~\cite{solar},
accelerator~\cite{k2k} and reactor~\cite{KAMLAND} neutrino data. If we
do not consider the claimed evidence for oscillations by the LSND
experiment~\cite{Athanassopoulos:1998pv}, that must be confirmed or
excluded by the ongoing MiniBooNE experiment~\cite{miniboone},
oscillations in the leptonic sector can be accommodated in the three
family Pontecorvo-Maki-Nakagawa-Sakata (PMNS) mixing matrix $U_{PMNS}$

\begin{eqnarray*}
\centering
\label{pmns}
U_\mathit{PMNS} =
\left (
\begin{array}{ccc}
c_{12}c_{13} & s_{12}c_{13} & s_{13}e^{-i\delta} \\ & & \\
-s_{12}c_{23}-\ \hfill& c_{12}c_{23}- \hfill & s_{23}c_{13} \\
\hfill \ \ c_{12}s_{23}s_{13}e^{i\delta} & \hfill \ \ s_{12}s_{23}s_{13}e^{i\delta} & \\
& & \\ s_{12}s_{23}-\hfill& -c_{12}s_{23}-\hfill & c_{23}c_{13} \\
\hfill \ \ c_{12}c_{23}s_{13}e^{i\delta} & \hfill \ \ s_{12}c_{23}s_{13}e^{i\delta} & \\
\end{array}
\right )
\end{eqnarray*}

where the short-form notation
$s_{ij}\equiv\sin\theta_{ij},c_{ij}\equiv \cos \theta_{ij}$ is used.
Further Majorana phases have not been introduced, since oscillation
experiments are only sensitive to the two neutrino mass squared
differences $\Delta m^2_{12}, \Delta m^2_{23}$ and to the four
parameters in the mixing matrix of Eq.~(\ref{pmns}): three angles and
the Dirac CP-violating phase, $\delta$.

There are several global fits of all available data. As an example we
report the ${\pm}\sigma$ ranges (95\%) as obtained in
Ref.~\cite{foglireview}:

$$\sin^2\theta_{13} = 0.9^{+2.3}_{-0.9}{\times}10^{-2}$$
$$\Delta m^2\theta_{12} = 7.92\pm0.09{\times}10^{-5}~\mbox{eV}^2$$
$$\sin^2\theta_{12} = 0.314^{+0.18}_{-0.15}$$
$$\Delta m^2\theta_{23} = 2.4^{+0.21}_{-0.26}{\times}10^{-3}~\mbox{eV}^2\,.$$

The next steps on the way of a full understanding
of neutrino oscillations by using neutrino beams produced at
accelerators are

\begin{itemize}
\item confirm the source of atmospheric neutrino oscillations,
i. e. observe the oscillation $\nu_\mu\rightarrow\nu_\tau$;
\item measure the remaining parameters of the PMNS mixing matrix:
$\theta_{13}$ and $\delta$;
\item measure the sign of $\Delta m^2_{23}$;
\item perform precision measurements of the angles
$\theta_{12}$ and $\theta_{23}$, and of
$\Delta m^2_{12}$ and $\Delta m^2_{23}$.
\end{itemize}

It is worth noting that there are other searches (like $\beta$-decay
and double-$\beta$ decay experiments, and space experiments studying
anisotropies in cosmic background radiation) which provide very
important information like the absolute value of the neutrino mass or
whether the neutrino is a Dirac or a Majorana particle. For a
comprehensive review of the analysis of these experiments we refer
to~\cite{foglireview} and references therein.

Among the oscillation parameters, a relevant role is played by the
mixing angle $\theta_{13}$. Indeed, as discussed in
Ref.~\cite{Apollonio:2002en,physrep}, a vanishing or too small value
for $\theta_{13}$, would make impossible the observation of the CP
violation in the leptonic sector and fix the neutrino mass hierarchy
(sign of $\Delta m^2_{23}$ exploiting matter effects). If
$\theta_{13}$ is large enough ($>3^\circ$) to allow for its discovery
by the forthcoming experiments~\cite{phaseIexps} (Phase I
experiments), new facilities and new
experiments~\cite{Apollonio:2002en,physrep} (Phase II experiments)
would be needed in order to precisely measure the PMNS matrix.

Several projects have been proposed for the Phase II (see
Ref.~\cite{Apollonio:2002en,physrep} and references therein). In this
paper we investigate a possible window of opportunity for the neutrino
oscillation physics compatible with the upgrade of the LHC (after
2015) that fully exploits european infrastructures and that has an
adequate sensitivity to the 1-3 sector of the PMNS matrix.

This paper is organized as follow. After a review of the proposed
neutrino beams in Europe~\ref{beams}, we focus on the $\beta$-beam
concept and in particular on a $\beta$-beam set-up based on the
so-called Super-SPS. We then discuss the proposed detector to exploit
the neutrino beam and finally present its physics reach.

\section{Neutrino Beams}
\label{beams}
Current neutrino oscillation experiments are based on beams where
neutrinos come from the decay of mesons produced in the interaction of
high energy protons impinging onto a target (typically Be or
graphite). However, such conventional beams have some limitations that
could be overcome by using new beam-line concepts: $\beta$-beams or
Neutrino Factories. For a comprehensive discussion of future beams and
their comparison we refer to~\cite{Apollonio:2002en,physrep} and
references therein.

\subsection{Conventional Neutrino Beams}
\label{conventional}

One can identify the main components of a conventional neutrino beam
line at a high energy accelerator as

\begin{itemize}
\item the target onto which protons are sent to produce pions and kaons;
\item the focusing system which guides the mesons
  along the desired neutrino beam direction;
\item the decay tunnel (usually evacuated) where mesons decay
  and produce neutrinos and muons.
\end{itemize}

From meson decay kinematics it follows that the neutrino energy is given by

\begin{equation}
E_\nu = {\frac{m^2_{\pi(K)}-m^2_\mu}{m^2_{\pi(K)}}}\frac{E_{\pi(K)}}{(1+\gamma^2
\theta^2)}
\label{enu_pi_K}
\end{equation}

where $\gamma$ is the Lorentz boost of the parent meson, $E_{\pi(K)}$
its energy and $\theta$ the angle of the neutrino with respect to the
meson flight direction.

There are three types of conventional neutrino beams: the Wide Band
Beams (WBB), the Narrow Band Beams (NBB) and the Off-Axis Beams
(OAB). WBB are characterized by a wide energy spectrum (they could
spread over a couple of order of magnitude) and correspondingly high
neutrino flux. Given these features, WBB are the optimal solution to
make discoveries. The drawback is that, if the signal comes from a
small part of the energy spectrum, it could be overwhelmed by the
background also induced by neutrinos outside the signal
region. Conversely, NBB may produce almost monochromatic energy
spectra. This can be obtained by selecting a small momentum bite of
the parent $\pi$ and $K$. However, the neutrino yield is significantly
reduced. This is an important drawback for oscillation searches.

A good compromise between the requirements of a high flux and a narrow
energy spectrum is obtained by means of Off-Axis
Beams~\cite{Para:2001cu}. This technique involves designing a
beam-line which can produce and focus a wide range of mesons in a
given direction (as in the WBB case), but then putting the detectors
at an angle with respect to that direction. Since the pion decay is a
two-body decay, a given angle between the pion direction and the
detector location corresponds to a given neutrino energy (almost)
independently of the pion energy. Furthermore, the smaller fraction of
high energy tails reduces the background from neutral-current (NC)
events, which can be misidentified for a $\nu_e$ charged-current (CC)
interaction due to the early showering of gamma's from the $\pi^0$
decay.

It is worth noting that, independent of the adopted solution, there
are common problems to all conventional neutrino beams

\begin{itemize}
\item the hadron yield in the proton-target interaction has large
uncertainties due to lack of data and to theoretical difficulties in
describing hadronic processes. This implies difficulties in predicting
the neutrino flux and spectrum with good accuracy;
\item in addition to the dominant flavor in the beam (typically
$\nu_\mu$) there is a contamination (at the few percent level) from
other flavors ($\bar{\nu}_\mu\,,\nu_e~\mbox{and}~\bar{\nu}_e $).
\end{itemize}

The knowledge of the beam spectrum and composition has a strong impact
both on the precision measurements of the angle $\theta_{23}$, on the
mass squared difference $\Delta m^2_{23}$ and on the sensitivity to
the mixing angle $\theta_{13}$. For instance, from the CHOOZ limit on
$\theta_{13}$ we know that the $\nu_\mu\rightarrow\nu_e$ appearance
probability is smaller than 5\%, which is of the same order of
magnitude of the beam contamination. Therefore, the observation of
$\nu_e$ appearance and the related $\theta_{13}$ measurement are
experimentally hard. Consequently, the usage of a close detector, to
solve the experimental problem related to the knowledge of the beam,
is mandatory.

In the last years a new concept of conventional beam (the so-called
``Super-Beam'') has been put forward in order to maximize the
sensitivity to $\theta_{13}$~\cite{Richter:2000pu}. Super-Beams will
provide a much higher neutrino flux, but at low energy (below 1 GeV).
This will open the possibility to perform long-baseline experiments
with high statistics and tuned at the oscillation maximum even at
moderate distances between source and detector.

\begin{figure*}[tbph]
\centering
\includegraphics[width=75mm]{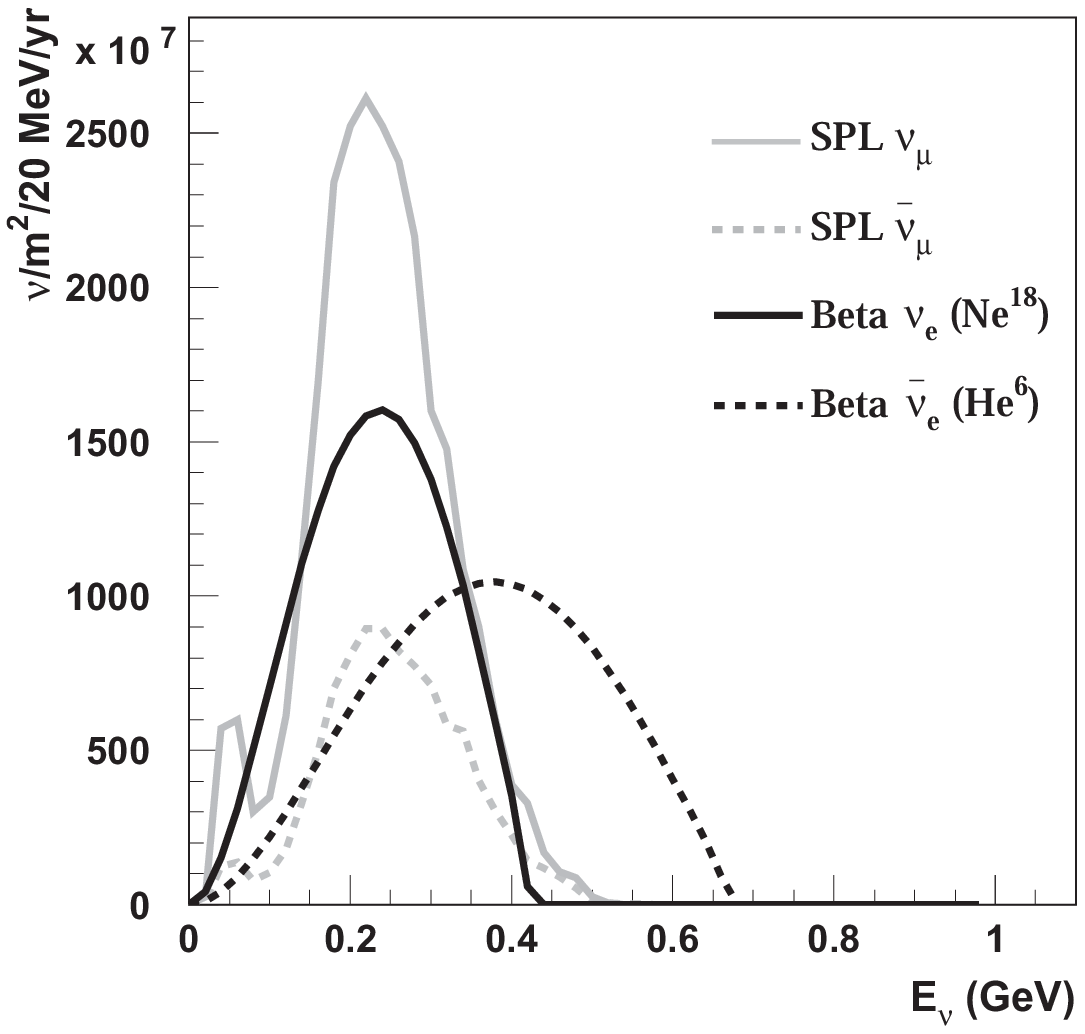}\includegraphics[width=75mm]{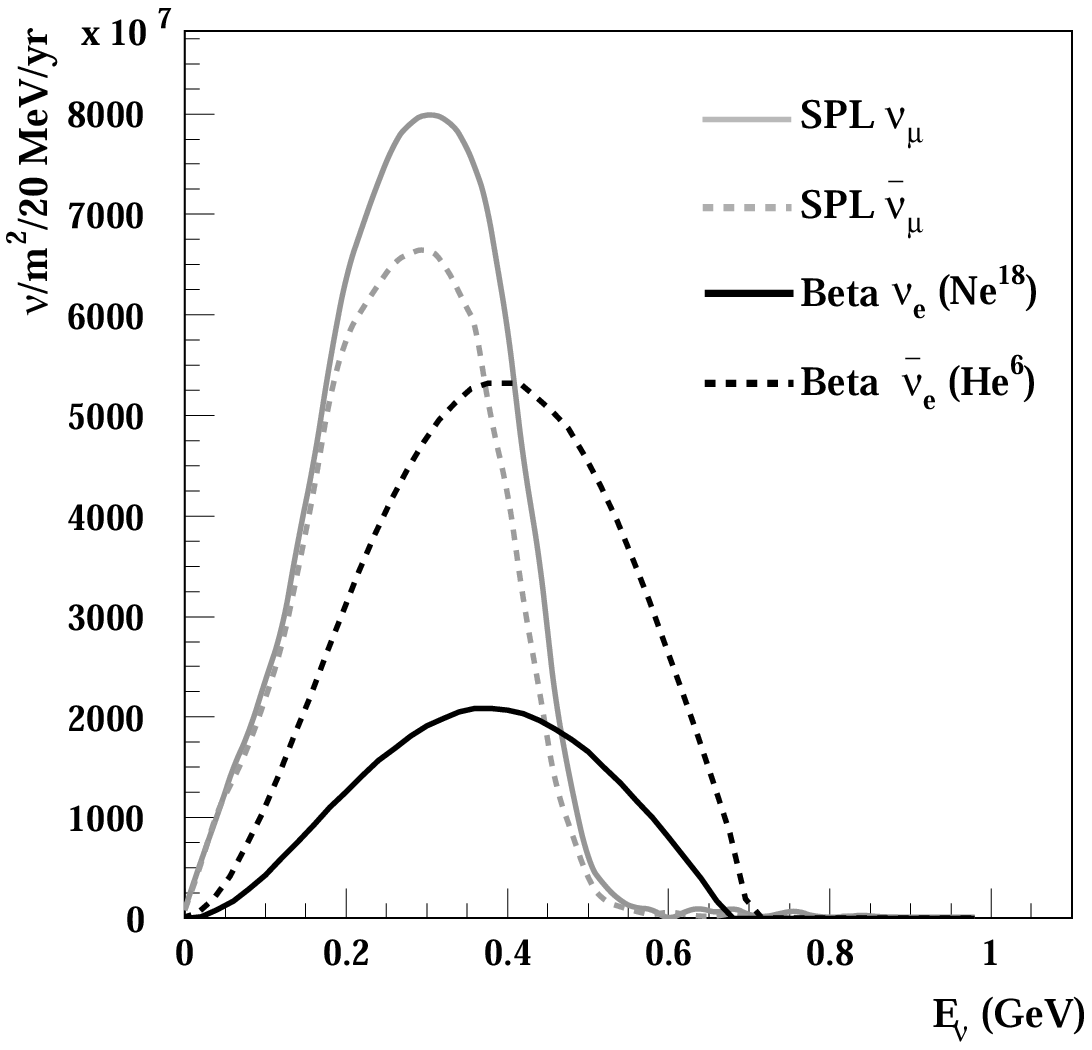}
\caption{Left: neutrino flux of $\beta$-Beam ($\gamma_{\He}=60$,
$\gamma_{\Ne}=100$, shared mode) and CERN-SPL SuperBeam,
           2.2 GeV, at 130 Km of distance.
           Right: the same for $\gamma_{\He}=100$,   $\gamma_{\Ne}=100$,
           (non shared mode, that is just one ion circulating in the decay
           ring)
           and a 3.5 GeV SPL Super-Beam.}
  \label{fig:fluxes}
 \end{figure*}

In Europe a 4~MW Superconducting Proton Linac (SPL)~\cite{SPL}, to be
built from scratch, would deliver a proton beam (with energy in the
range 2.2~GeV~\cite{Garoby:2005se}-3.5~GeV~\cite{Campagne:2004wt}) on
a Hg target to generate an intense (anti-)neutrino flux from the
$\pi^+$ ($\pi^-$) decay. This intense neutrino beam, whose fluxes are
shown in Fig.~\ref{fig:fluxes}, has been proposed to be sent from CERN
towards the Frejus Laboratory. The average energy of neutrinos
produced with this facility is of the order of few hundred MeV. The
physics potential and the accelerator complex needed for such a
Super-Beam are discussed in Refs.~\cite{physrep} and references
therein. Here, we only recall that a SPL based neutrino program will
improve by about one order of magnitude the T2K $\theta_{13}$
sensitivity, while it will be able to address neither the CP violation
in the leptonic sector nor the neutrino mass hierarchy.

The SPL is the not a mandatory solution for the energy/luminosity
upgrade of the LHC~\cite{garobyarci}, while it is an essential
component of a Neutrino Factory complex~\cite{Apollonio:2002en}. The
low energy of neutrinos produced with this facility has an important
drawback on the detector choice. Indeed, in order to compensate the
small cross-section and to allow an efficient particle identification
huge and low density detectors are mandatory. The typical detector
proposed to exploit a neutrino beam from a SPL is a Megaton water
Cerenkov detector~\cite{physrep}. The proposed location for this
detector is the Frejus laboratory, where a cavern capable to host it
should be built. All in all, we think that an experimental program
based on a SPL is uprooted with respect to a possible common effort of
the elementary particle community in Europe. Moreover, it assumes the
construction of a million cubic meter cavern capable to host the
detector. Last but not least, the neutrino beam has no special
advantages (the main sources of systematics are still there) with
respect to the existing ones, but the higher intensity. Conversely,
being a low energy beam one has to deal with a region where Fermi
motion has an energy comparable with the one of the incident neutrino
beam. Consequently it is not possible to measure the neutrino energy
spectrum, but only count neutrinos of a given flavour.

\subsection{Neutrino Factory}
\label{nufact}

\begin{figure*}[tbph]
\centering
\includegraphics[width=100mm]{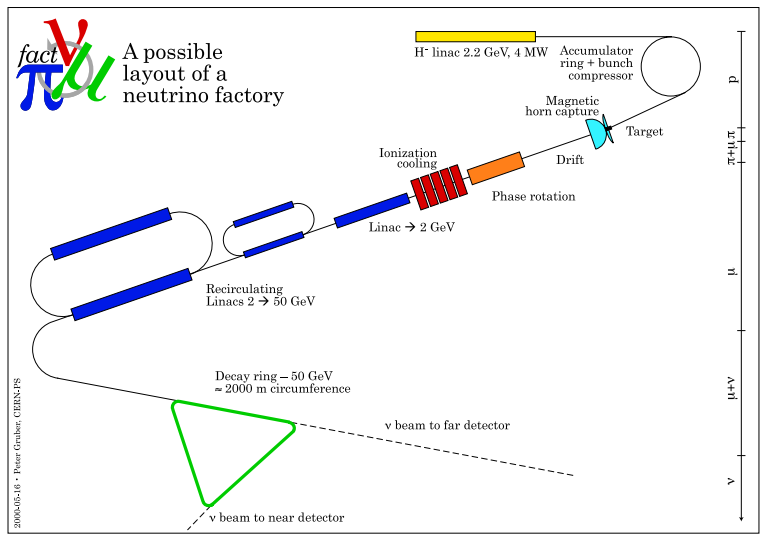}
\caption{Expected layout for a Neutrino Factory at CERN.}
  \label{fig:nufact}
 \end{figure*}

The first stage of a Neutrino Factory is similar to that of a
Super-Beam. Namely, protons are sent onto a target producing pions and
kaons that are collected by means of magnetic lenses. However, while
in those beams hadrons are let decay launching neutrinos toward the
detector site, in a Neutrino Factory daughter muons are collected and
accelerated in a ring with long straight sections. Muon decays in each
straight section generate highly collimated neutrino beams. The
expected layout for a Neutrino Factory at CERN is shown in
Fig.~\ref{fig:nufact}. If $\mu^+$ are stored, $\mu^+\rightarrow
e^+\nu_e\bar{\nu}_\mu$ decays generate a beam consisting of 50\%
$\nu_e$ and 50\% $\bar{\nu}_\mu$. Similarly, if $\mu^-$ are stored the
beam consists of 50\% $\nu_\mu$ and 50\% $\bar{\nu}_e$. Since the
kinematics of muon decay is well known, we expect minimal systematic
uncertainties on the neutrino flux and spectrum. Hence, compared to
conventional neutrino beams, Neutrino Factories provide $\nu_e$ and
$\bar{\nu}_\mu$ beams or $\nu_\mu$ and $\bar{\nu}_e$ beams, with small
systematic uncertainties on the flux and spectrum. Radiative effects
on the muon decay have been calculated and amount to about $4{\times}10^{-3}$
with a much smaller error. Overall, the flux is expected to be known
with a precision of the order of $10^{-3}$. Another important feature
of a Neutrino Factory beam is its sharp cut-off at the energy of the
stored muons. In a conventional neutrino beam there is a high-energy
tail which, as already mentioned, gives rise to background from NC
events in which a leading $\pi^0$ is mis-interpreted as an electron,
faking $\nu_\mu\rightarrow\nu_e$ signal.  Furthermore, the possibility
to store high-energy muons that in turn produce high-energy neutrinos
opens the study of oscillation channels like
$\nu_\mu\rightarrow\nu_\tau$ and $\nu_e\rightarrow\nu_\tau$, whose
combined physics potential has been discussed in~\cite{donini}.

Summarizing, the Neutrino Factory provides an excellent neutrino beam
optimal for both neutrino oscillation searches and other
physics~\cite{Apollonio:2002en}. However, it is based on a very
challenging technology and it has no relevant overlap with the present
(and future) CERN accelerators. As far as the exploitation of existing
infrastructure, while there is no need for a large cavern to host a
megaton detector, the Gran Sasso halls could be too close, if the
Neutrino Factory is built at CERN. Consequently, a new underground
laboratory has to be built in order to host the far detector. It is
worth noting that given the very high intensity and high energy of a
Neutrino Factory the far detector will have a reasonable size
($\mathcal{O}(10^4\,m^3)$ not $\mathcal{O}(10^6\,m^3)$).

\subsection{$\beta$-beams}
\label{betabeam}

A $\beta$-beam~\cite{Zucchelli:sa} is made by accelerating radioactive
ions with a short beta-decay lifetime, by storing them in a ring with
straight sections and by letting them decay. The focusing of the beam
is provided by the Lorentz boost. Having the possibility to accelerate
either $\beta^-$ (e.g. $^6$He) or $\beta^+$ (e.g. $^{18}$Ne) ions,
pure $\bar{\nu}_e$ or pure ${\nu}_e$ beams can be produced,
respectively. In order to illustrate the value of the $\beta$-beam
concept, we briefly discuss the production of an anti-neutrino beam. A
good beta-emitter for anti-neutrino production is the
$^6\mbox{He}^{++}$ ion that decays into
$^6_3\mbox{Li}^{++}e^-\bar{\nu}_e$ with a $\beta$-decay endpoint
($E_0$) of about 3.5~MeV. The anti-neutrino spectrum is precisely
known from laboratory measurements of the associated electron, since
$E_e + E_\nu \approx E_0$. Since the ion is spin-less, decays at rest
are isotropic. When ions are accelerated ($\gamma$ values up to 150
are possible) the neutrino transverse momentum in the laboratory frame
is identical to that observed in the rest frame, while the
longitudinal momentum is multiplied by a factor $\gamma$. Therefore,
neutrino beam divergence is of the order of $1/\gamma$ (less than
10~mrad for $\gamma=100$), and the average neutrino energy in the
forward direction is $2\gamma\mbox{E}_{cms}\sim$500~MeV.

The technical feasibility of accelerating ions, although at relatively
low energies, has been already demonstrated in nuclear physics
experiments such as at ISOLDE at CERN. Given the small neutrino
energy, a potential drawback of this approach is the substantial
background from atmospheric neutrinos. To overcome this problem, ion
beams should be bunched. At present, this is a major technical issue.

In the baseline design, the proton driver for a $\beta$-beam is the
proposed SPL~\cite{SPL}. However, contrary to naive expectation, a
multi-megawatt booster is not necessary for the construction of a beta
beam or a nuclear physics (EURISOL-like~\cite{betabeams_moriond}).
Indeed, independently of the $\gamma$, a $\beta$-beam requires a
$\sim200$~kW proton driver operating in the few GeV region. The
collection and ionization of the ions is performed using the ECR
technique. Hereafter ions are bunched, accelerated and injected up to
the high energy boosters.

Summarizing, the main features of a neutrino beam based on the
$\beta$-beam concept are:

\begin{itemize}
\item the beam energy depends on the $\gamma$ factor. The ion
accelerator can be tuned to optimize the sensitivity of the
experiment;
\item the neutrino beam contains a single flavor with an energy
spectrum and intensity known a priori. Therefore, unlike conventional
neutrino beams, close detectors are not necessary to normalize the fluxes;
\item neutrino and anti-neutrino beams can be produced with a
comparable flux;
\item
Differently from Super-Beams, $\beta$-beams experiments search for
$\nu_e
\rightarrow \nu_\mu$ transitions, requiring a detector capable to
identify muons from electrons.
Moreover, since
the beam does not contain $\nu_\mu$ or $\bar{ \nu}_\mu$ in the initial
state, magnetized detectors are not needed. This is in contrast with
the neutrino factories (see below) where the determination of the muon
sign is mandatory.
\end{itemize}

A baseline study for  a $\beta$-beam complex has been carried out at
CERN~\cite{Lindroos}. The SPS could accelerate \He\ ions at a maximum
$\gamma$ value of $\gamma_{\He}=150$ and \Ne\ ions up to
$\gamma_{\Ne}=250$. In this scenario the two ions circulate in the
decay ring at the same time. A feasible option provided that their
$\gamma$ are in the ratio $\gamma_{\He}/\gamma_{\Ne}=3/5$.
The reference $\beta$-beam fluxes  are $2.9{\times}10^{18}$ \He\ useful
decays/year and $1.1{\times}10^{18}$ \Ne\  decays/year if the two ions are
run at the same time in the complex. Novel developments, suggesting
the possibilities of running the two ions separately at their optimal
$\gamma$~\cite{MatsPrivate}, have recently triggered a new optimal
scheme for the $\beta$-beam. In this scheme both ions are accelerated
at $\gamma=100$. The expected fluxes for the baseline scenario are
shown in Fig.~\ref{fig:fluxes}.

$\beta$-beam capabilities for ions accelerated at higher energies than
those allowed by SPS have been computed in
\cite{latestJJ,HighEnergy,HighEnergy2}. In the next Section we focus
on a possible accelerator complex needed to build a $\beta$-beam with
$\gamma$ in the range 350-580 and we compare it to the baseline
design. We refer in the following to this set-up as the medium
$\gamma$ scenario.

\section{The medium $\gamma$ scenario}

\subsection{The accelerator complex}

\begin{figure}[tb]
\centering
\includegraphics[width=75mm]{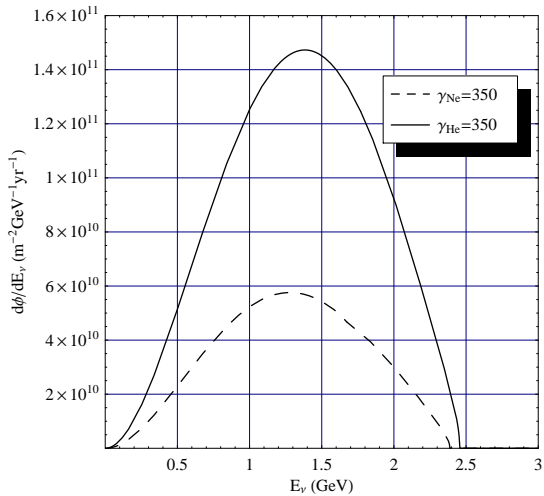}
\caption{$\beta$-Beam fluxes at the Gran Sasso location (735 km baseline) as a function of
the neutrino energy for $\gamma = 350$.}
\label{fig:fluxes2}
\end{figure}

The choices and timescale for the upgrades of the LHC will depend on
the feedbacks from the first years of data taking. Still, three phases
can already be envisaged~\cite{bruning,scandale}: an optimization of
present hardware (``phase 0'') to reach the ultimate luminosity of $2{\times}
10^{34}~\mbox{cm}^{-2}~\mbox{s}^{-1}$ at two interaction points; an
upgrade of the LHC insertions (``phase 1'') and, finally, a major
hardware modification (``phase 2'') to operate the LHC in the ${\cal
L} \simeq 10^{35}~\mbox{cm}^{-2}~\mbox{s}^{-1}$ regime and, if needed,
prepare for an energy upgrade. The most straightforward approach to
``phase 2'' would be the equipment of the SPS with fast cycling
superconducting magnets in order to inject protons into the LHC with
energies of about 1~TeV.
The 1~TeV injection option (``Super-SPS'') would have an enormous
impact on the design of a $\beta$-beam at CERN. This machine fulfills
simultaneously the two most relevant requirements for a high energy
$\beta$-beam booster: it provides a fast ramp ($dB/dt=1.2 \div
1.5$~T/s~\cite{fabbricatore}) to minimize the number of decays during
the acceleration phase and, as noted in ref.\cite{HighEnergy}, it is
able to bring \He\ up to $\gamma\simeq 350$ (\Ne\ to $\gamma
\simeq 580$)\footnote{It is worth mentioning that the Super-SPS eases
substantially injection of $\beta$ ions in the LHC to reach $\gamma \gg
350-580$. For a discussion of this option we refer
to~\cite{HighEnergy2}.}. The mean neutrino energies of the
$\bar{\nu}_e$, ${\nu}_e$ beams corresponding to a $\gamma = 350$ are
1.36~GeV and 1.29~GeV, respectively. Fig.~\ref{fig:fluxes2} shows the
$\beta$-Beam neutrino fluxes computed at the 735 Km baseline, keeping
$m_e \neq 0$ as in Ref.~\cite{HighEnergy}.

\begin{figure*}[tb]
\centering
\includegraphics[width=140mm]{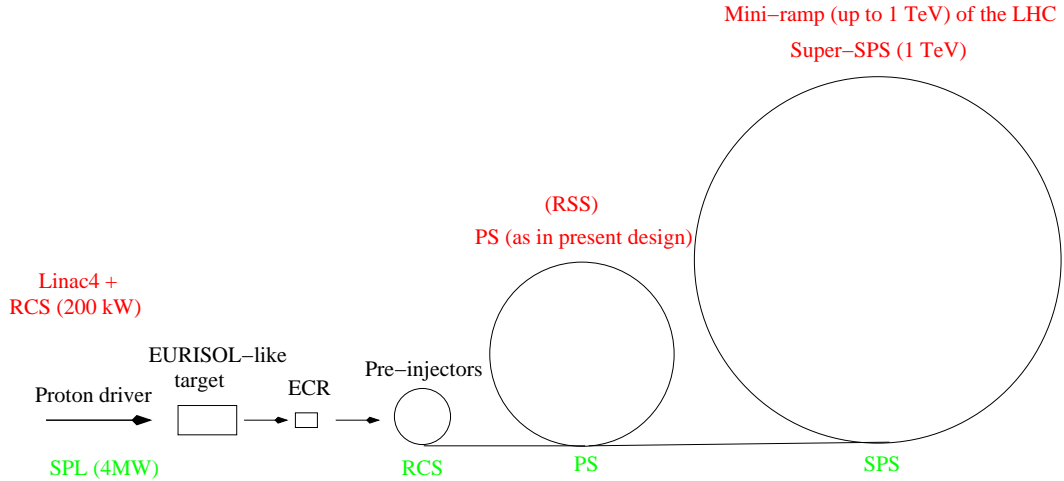}
\caption{The main component of the $\beta$-beam complex up to
injection into the decay ring~\cite{inpreparation}. In the lower part,
the machines considered in the baseline option are indicated. The
alternatives that profit of the upgrade of the LHC injection system
are also mentioned (upper part).  RCS stands for Rapid Cycling
Syncrotron, RSS for Rapid Superconducting
Syncrotron~\cite{garoby_hif04}.}
\label{fig:complex}
\end{figure*}

The increase of the ion energy in the last element of the booster
chain represents a challenge for stacking~\cite{terranova_nufact04}.
Ions of high rigidity must be collected in a dedicated ring of
reasonable size. In the baseline design, this is achieved by a decay
ring made of small curved sections (radius $R\sim 300$~m) followed by
long straigth sections ($L=2500$~m) pointing toward the far neutrino
detector.  In this case, the decays that provide useful neutrinos are
the ones occurring in the straigth session where neutrinos fly in the
direction of the detector and the useful fraction of decays
(``livetime'') is limited by the decays in the opposite arm of the
tunnel. For the CERN to Frejus design the livetime is $L/(2\pi R+2L)
\sim 36$\% and the overall length has been fixed to 6880~m. A decay
ring of the same length equipped with LHC dipolar magnets (8.3~T)
would stack ion at the nominal Super-SPS rigidity with a significantly
larger radius ($\sim 600$~m). The corresponding lifetime is thus 23\%.
The actual intensities that can be achieved with a high energy booster
and possible losses with respect to the baseline design still need a
dedicated machine study\footnote{for recent progresses in the
framework of the baseline design (SPS-based),
see~\cite{nota_lindroos}.}. Hence in the following, physics
performances are determined as a function of fluxes. Nevertheless, the
looser constrains on the time structure of the beam and the occupancy
of the decay ring~\cite{inpreparation} offers a way out for
compensation of the losses due to the decrease of the number of decay
per unit time.
We remind that the baseline
$\beta$-beam design aims at $2.9{\times} 10^{18}$~\He\ and
$1.1{\times} 10^{18}~$\Ne\ decays per year.  Fig.~\ref{fig:complex}
sketches the main components of the $\beta$-beam complex up to
injection into the decay ring. In the lower part, the machines
considered in the baseline option are listed. The alternatives that
profit of the upgrade of the LHC injection system are also mentioned
(upper part). For a review of technical challenges of $\beta$-beams,
we refer to~\cite{matsnufact}.

\subsection{Baseline versus medium $\gamma$ scenario}

The comparison of the two scenarios can be performed at two different
levels: the first one considering the synergies with both the CERN
accelerator complex and the existing underground laboratories; the
second looking at their physics reach.

From the previous discussions it is straightforward that the baseline
scenario presents a strong synergy with the present CERN accelerator
complex. On the other hand, it could profit of an upgrade of the PS
machine. Indeed, at the present one of the main limitations comes from
the losses in the PS for He that make very difficult the maintenance
of the machine. Furthermore, it foresees the construction of a very
expensive SPL. Conversely, the medium $\gamma$ scenario fully exploits
the machine upgrades for the LHC energy/luminosity upgrade. At the
typical energies of the medium $\gamma$ scenario the peak of
oscillation probability is comparable to the CERN to Gran Sasso
distance. Therefore, it would leverage the existing infrastructure at
the Gran Sasso Laboratories as possible site for the far detector.
Furthermore, due to the large increase of cross-section the use of
dense detectors (reduction of the detector mass/volume) would be
possible compared with the baseline design. The latter foresees a
Megaton water Cerenkov detector to be installed in an underground
laboratory that should be built from scratch at the Frejus site.

One of the problems of the baseline scenario is the measurement of the
$\nu_\mu$ and $\bar{\nu}_\mu$ cross-sections. Indeed, a near detector
may measure with high accuracy the $\nu_e$ and $\bar{\nu}_e$
cross-sections, but the signal ones. On the other hand, given the
smallness of the neutrino energy, the mass difference between the
electron and the muon starts to be important. The medium $\gamma$ has
another advantage with respect to the baseline scenario. Indeed, while
there are already planned experiments aiming at the few percent
precision cross-section measurement for both  $\nu_\mu$ and
$\bar{\nu}_\mu$ in the 1 GeV region, there are no plans for such a
measurements in the few hundred MeV region. Therefore, in the baseline
scenario one plans to use the SPL beam to measure the $\nu_\mu$ and
$\bar{\nu}_\mu$ cross-sections, while it is not so important for the
oscillation measurements~\cite{Donini:2004hu}. The impact of the
systematic error on the CP-violation discovery potential has been
studied in Ref.~\cite{mauronufact}



\section{The detector at the Gran Sasso and its performances}

As already pointed out in the previous Sections, the main advantage of
working with a $\beta$-beam is that there is no need of a magnetized
detector to discriminate among neutrinos and anti-neutrinos. The only
requirement is a good muon identification in order to observe
$\nu_\mu$ or $\bar{\nu}_\mu$ coming from $\nu_e$ or $\bar{\nu}_e$
oscillations, respectively. Consequently, working at high (above
1~GeV) neutrino energies opens the possibility to exploit
non-magnetized iron calorimeters, i.e. high density detectors that can
operate beyond the ``single ring'' region of Water Cherenkov and that
can be hosted in relatively small underground sites. On top of a good
muon identification, these detectors also guarantee the energy
measurement of the hadronic shower produced in the neutrino
interaction. The measurement of the muon momentum and of the hadronic
shower allows for the reconstruction of the incident neutrino energy.

\begin{figure*}[tbph]
\centering
\includegraphics[width=75mm]{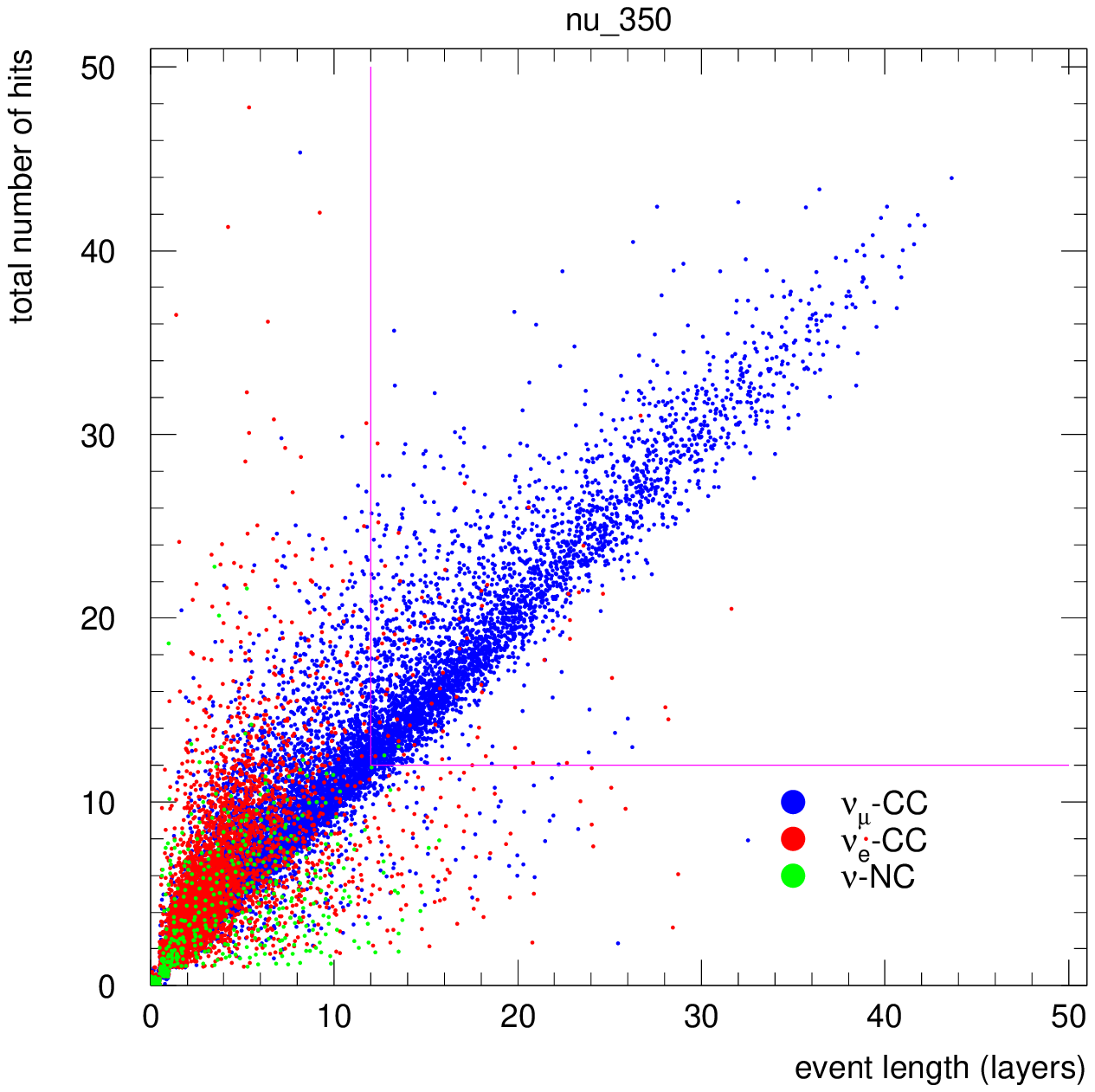}\includegraphics[width=75mm]{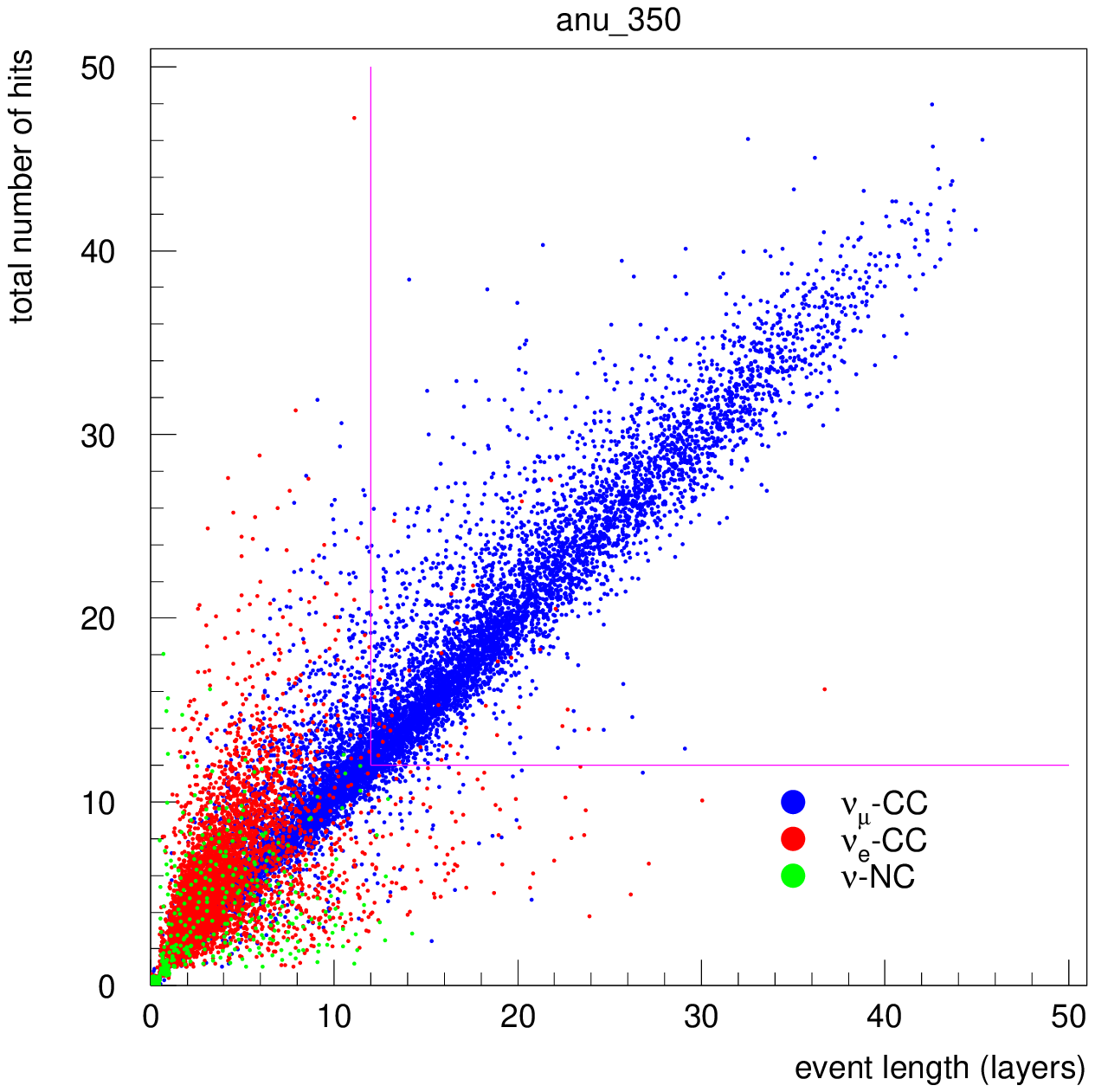}
\caption{Scatter plot of the total number of hits recorded in the detector versus
the total length (given in number of crossed layers) of the event for
neutrinos (left) and anti-neutrinos (right) with $\gamma=350$.
.}
  \label{fig:eveclass}
\end{figure*}

Several experimental techniques can be employed for the detector
design.  Among them, in~\cite{inpreparation} we considered a design
derived from a digital RPC based calorimeter proposed for the
reconstruction of the energy flow at the ILC detector~\cite{tdf}
(DHCAL). It consists of a sandwich of 4~cm non-magnetized iron and
glass RPC with an overall mass of 40kton. This detector could be
hosted in an underground site of LNGS. The active part of the RPC is
segmented in $2{\times}2~\mbox{cm}^2$ elementary cells. Details on the
detector structure and on the performance of the RPC's may be found in
Ref.~\cite{tdf}. A full Monte Carlo simulation of the DHCAL has been
implemented with the GEANT3 package and validated by comparing its
response with pion data with energy in the range from 2~GeV to
10~GeV~\cite{gustavino}. We used the full Monte Carlo simulation in
order to evaluate the detector response, but the event classification
capability is only based on inclusive variables (total number of hits,
event length (expressed in terms of number of crossed iron
layers)). The scatter plot of the event length versus the total number
of hits of the event is shown in Fig.~\ref{fig:eveclass} for neutrinos
(left panel) and anti-neutrinos (right panel) both coming from ions
accelerated at $\gamma= 350$. We plot all together $\nu_\mu$ and
$\nu_e$ charged-current (CC) interactions as well as neutrino
neutral-current (NC) interactions. We classify an interaction as a
$\nu_\mu$ CC-like event if both the event length and the total number
of hits in the detector are larger than 12. In the case the \Ne\ is
ran at $\gamma= 580$, we classify an event as a CC-like interaction if
the event length and the total number of hits are larger than 15 and
17, respectively. The typical efficiency for identifying a neutrino or
anti-neutrino CC interaction is, averaged out over the whole spectrum,
of the order of 50-60\%. Conversely, the probability for the
background to be identified as a CC-like event is smaller than 1\%.

\begin{figure*}[tbph]
\centering
\includegraphics[width=75mm]{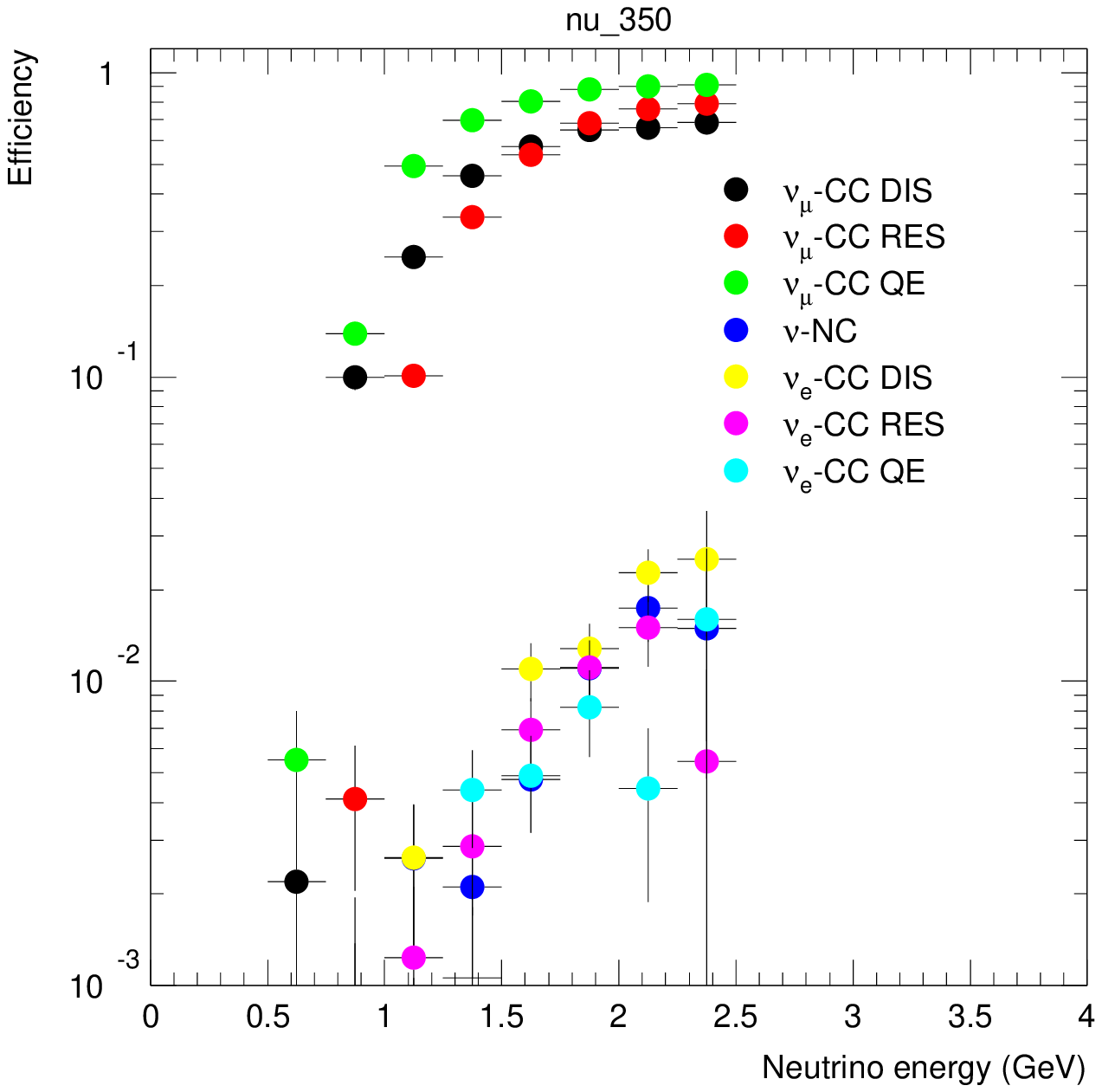}\includegraphics[width=75mm]{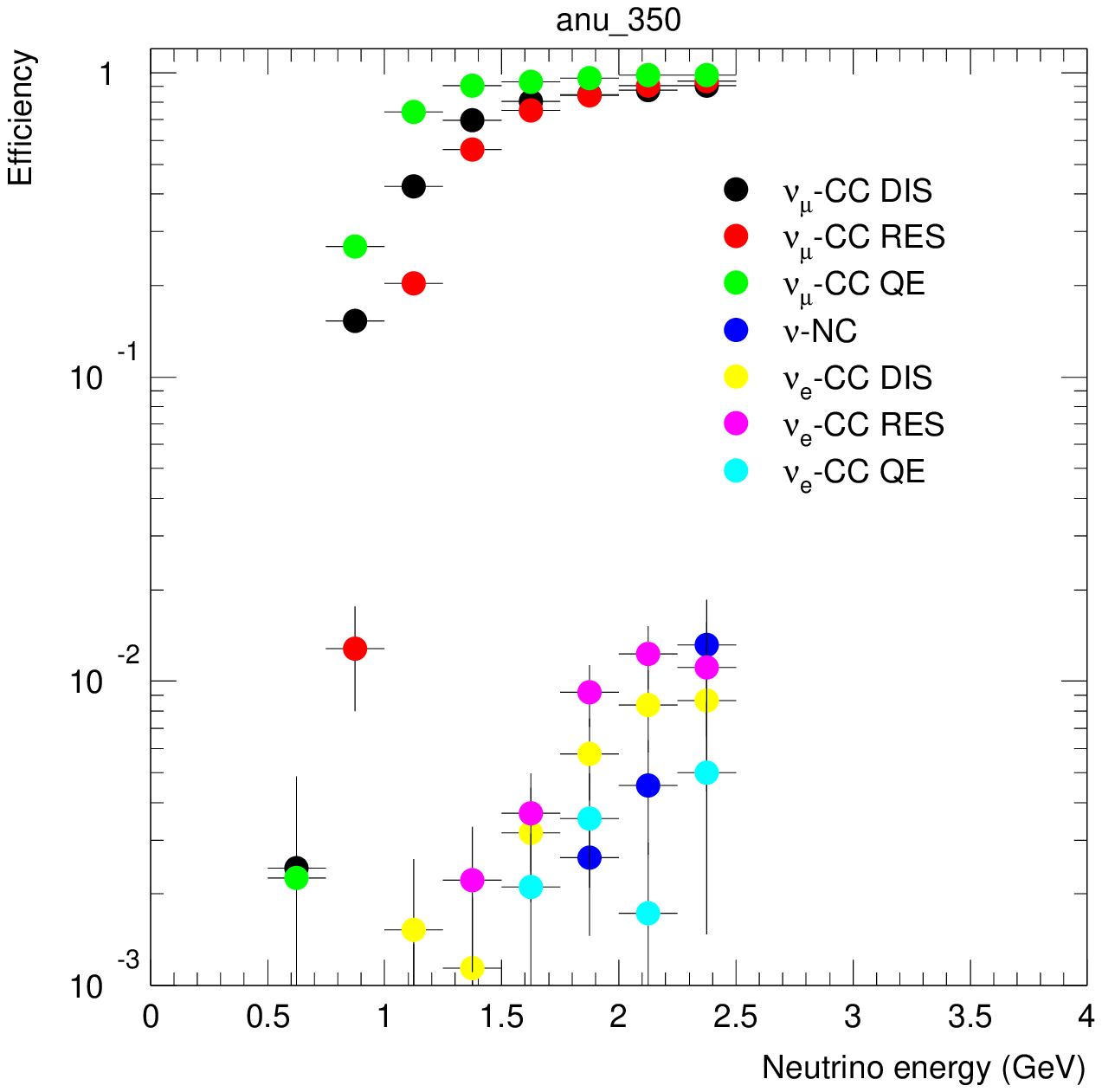}
\caption{Efficiencies for the signal ($\nu_\mu$ and $\bar{\nu}_\mu$
charged-current interactions) to be identified as CC-like event and
for the background ($\nu_e$ and $\bar{\nu}_e$ interactions, and
$\nu_\mu$ and $\bar{\nu}_\mu$ neutral-current interactions) to be
mis-identified as a CC-like events.}
  \label{fig:eveclasseff}
\end{figure*}

The efficiencies to correctly identify $\nu_\mu$
and $\bar{\nu}_\mu$ charge-current interactions are shown, separately
for deep-inelastic (DIS), quasi-elastic (QE) and resonance (RES)
production, in Fig.~\ref{fig:eveclasseff} as well as the probability
that $\nu_e$ and $\bar{\nu}_e$, separately for DIS, QE and RES
production, and neutral-current interactions are identified as a
CC-like interaction.

\section{Physics reach of the medium $\gamma$ scenario}
As discussed in the previous Sections, the expected neutrino
flux as a function of the $\gamma$ is still under evaluation.
Therefore, in the following we evaluate the physics reach as a
function of the flux normalized to the one assumed in the baseline
design ($F_0$). Finally, we note that in this work only the intrinsic
degeneracy is taken into account. For an exhaustive discussion on the
problem of the degeneracies at a $\beta$-beam complex, we refer
to~\cite{Donini:2004hu} and references therein.

The expected number of events as a function of $\delta$ and
$\theta_{13}$ has been obtained in the framework of a three family
scenario and it is shown in Table~\ref{event}. For the already
measured parameters, we assumed the following values: $\Delta
m^2\theta_{12} = 8{\times}10^{-5}~\mbox{eV}^2;\, \theta_{12} = 30^\circ;\,
\Delta m^2\theta_{23} = 2.5{\times}10^{-3}~\mbox{eV}^2$.

\begin{table}[hbt]
\begin{center}
\caption{Event rates for a 10 years exposure at a medium $\gamma$ $\beta$-beam of a 40 kton detector.
The observed oscillated charged-current events for different values of
$\delta$ and $\theta_{13}$, assuming the normal neutrino mass
hierarchy and $\theta_{23}=45^\circ$, are given. The expected
background is also reported.}
\begin{tabular}{|c|c|c|c|c|c|}
\hline
$\theta_{13}$ & $\delta$ & $\nu_\mu$CC & $\bar{\nu}_\mu$CC &
$\nu$-back. & $\bar{\nu}$-back.\\
\hline
$1^\circ$  & $-90^\circ$ & 1.45 & 37.67 & 126.02 & 77.28\\
\hline
$5^\circ$  & $-90^\circ$ & 103.46 & 271.05 & 126.02 & 77.28\\
\hline
$10^\circ$  & $-90^\circ$ & 532.23 & 863.18 & 126.02 & 77.28\\
\hline
$1^\circ$  &$0^\circ$ & 18.18 & 22.74 & 126.02 & 77.28\\
\hline
$5^\circ$  & $0^\circ$ & 187.02 & 196.49 & 126.02 & 77.28\\
\hline
$10^\circ$  & $0^\circ$ & 698.70 & 714.63 & 126.02 & 77.28\\
\hline
$1^\circ$  & $90^\circ$ & 32.04 & 2.57 & 126.02 & 77.28\\
\hline
$5^\circ$  & $90^\circ$ & 256.23 & 95.75 & 126.02 & 77.28\\
\hline
$10^\circ$  & $90^\circ$ & 836.60 & 513.91 & 126.02 & 77.28\\\hline
\end{tabular}
\label{event}
\end{center}
\end{table}

\subsection{Extraction of the neutrino oscillation parameters in presence of signal}

\begin{figure*}[tbhp]
\centering
\includegraphics[width=75mm]{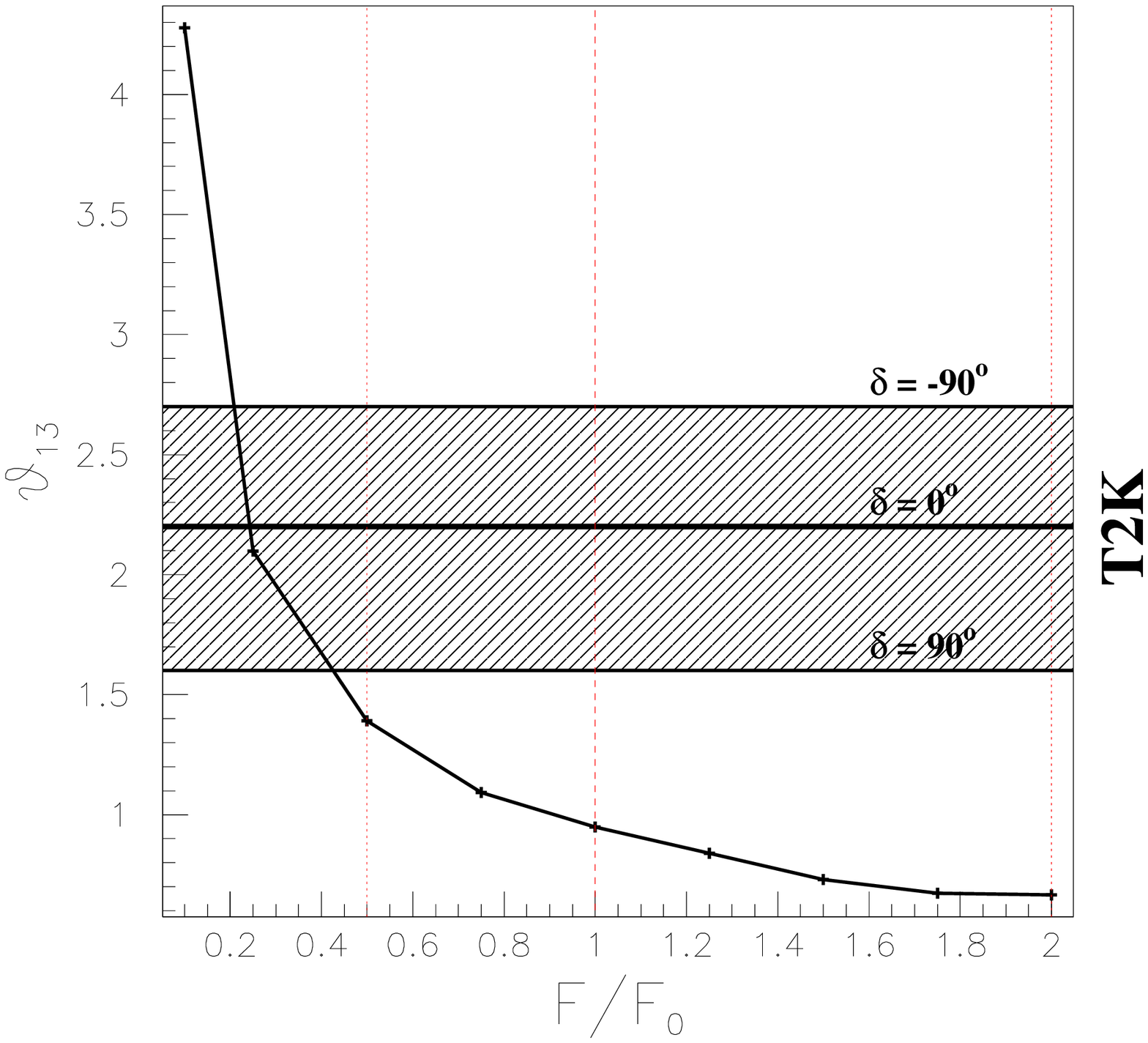}\includegraphics[width=75mm]{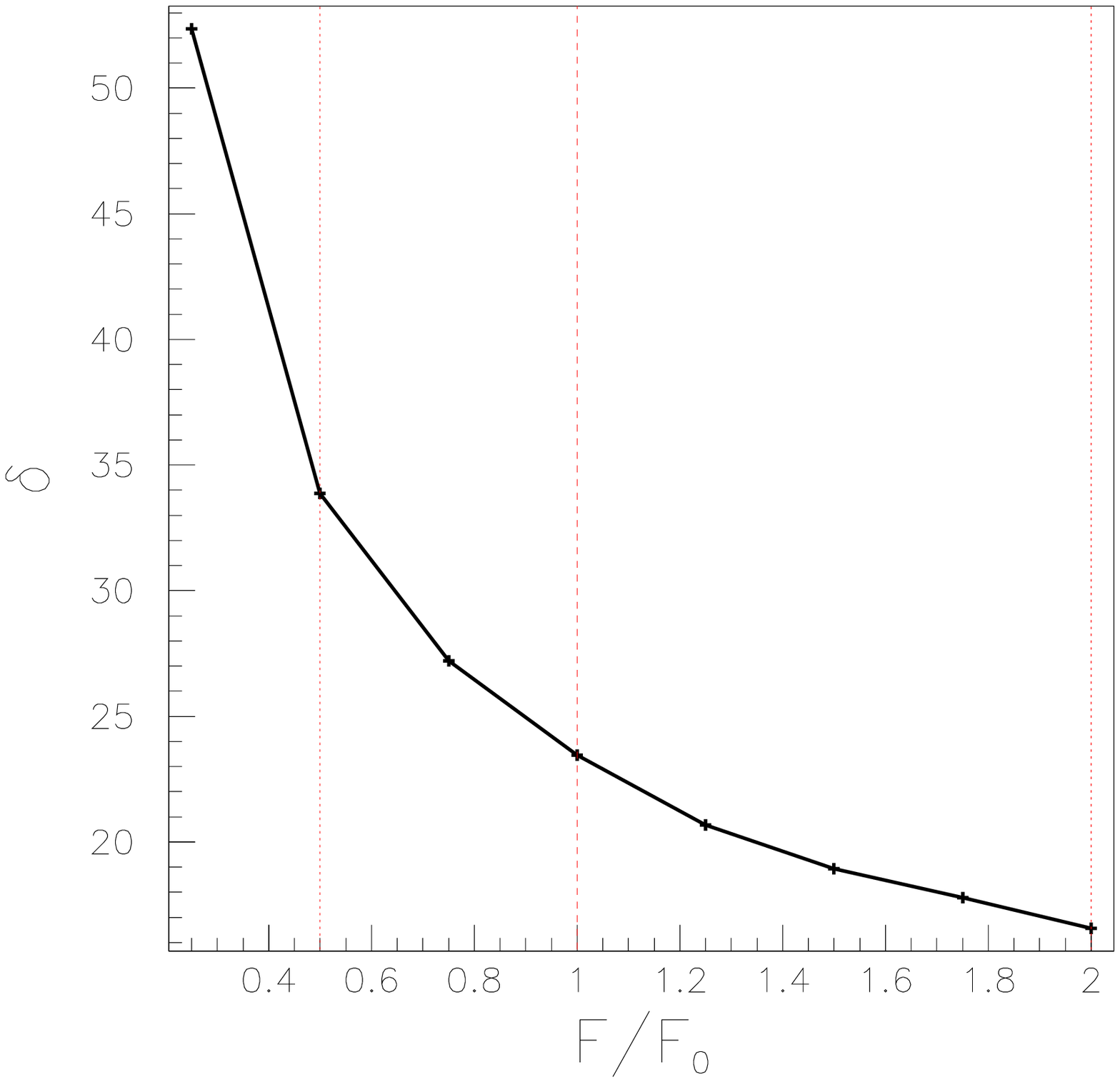}
\caption{Left plot: minimum $\theta_{13}$ that can be distinguished from zero at 99\%
C.L. as a function of the flux (1 corresponds to $F_0$). Right plot:
minimum $\delta$ that can be distinguished from zero, at 99\% C.L., as
a function of the neutrino flux.}
  \label{fluxstudy}
\end{figure*}

\begin{figure}[tbph]
\centering
\includegraphics[width=75mm]{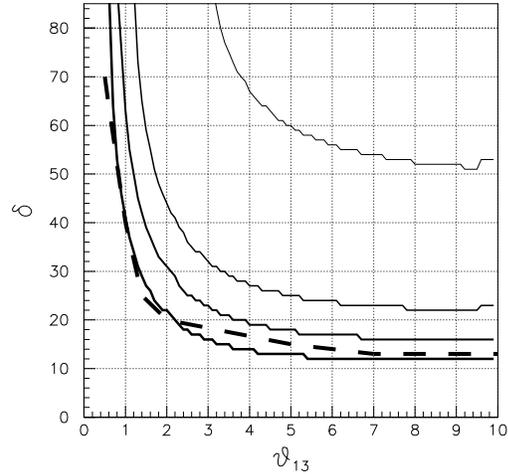}
\caption{$\delta_{CP}$ discovery potential at 99\% C.L. as a function of $\theta_{13}$.
The different solid lines corresponds at different fluxes. From left
to right: $2{\times}F_0$, $F_0$, $F_0/2$ and $F_0/10$. The dashed line show
the discovery potential for the baseline scenario as computed in
Ref.~\cite{mauronufact}.}
  \label{deltadisc}
\end{figure}

Since the neutrino flux from a $\beta$-beam is not yet well defined,
we plot in Fig.~\ref{fluxstudy} (left panel), for $\delta=90^\circ$,
the minimum $\theta_{13}$ that can be distinguished from zero at 99\%
C.L. as a function of the flux (1 corresponds to $F_0$). Notice that,
if the flux is at least half of $F_0$, it is possible to discover a
non vanishing $\theta_{13}$ even in the case of no signal observed in
the T2K experiment. Assuming a flux equal to $F_0$, values of
$\theta_{13}$ down to $1^\circ$ can be distinguished from zero.
Fig.~\ref{fluxstudy} (right panel) shows, for $\theta_{13}=3^\circ$,
the minimum $\delta$ that can be distinguished from zero, at 99\%
C.L., as a function of the neutrino flux. The value
$\theta_{13}=3^\circ$ has been chosen being the minimum value for
which T2K may discover a non-zero $\theta_{13}$. Also in this case,
unless the flux is smaller than $F_0/10$, it would be possible for the
whole $\theta_{13}$ range covered by the T2K discovery potential
discover CP violation in the leptonic sector. The minimum
$\delta_{CP}$ that can be discovered at 99\% C.L., as a function of
$\theta_{13}$, is shown in Fig.~\ref{deltadisc}~\footnote{A more
extensive analysis, exploiting particularly the energy resolution of
the detector is in progress~\cite{inpreparation}.}. As for comparison,
the discovery potential of the baseline scenario is also reported. We
can argue that, down to fluxes half of $F_0$ and for
$\theta_{13}>3^\circ$ (the discovery region of T2K), the discovery
potential of the baseline and of the medium $\gamma$ scenarios are
comparable.

\subsection{Exclusion plots in absence of signal}
\begin{figure}[tbph]
\centering
\includegraphics[width=75mm]{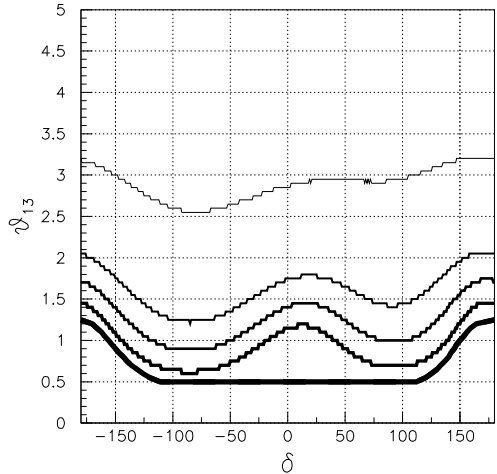}
\caption{$\theta_{13}$ discovery potential at 90\% C.L. as a function of $\delta$.
The different solid lines corresponds at different fluxes. From down
to top: $2{\times}F_0$, $F_0$, $F_0/2$ and $F_0/10$. The dashed line show the
discovery potential for the baseline scenario as computed in
Ref.~\cite{mauronufact}.}
  \label{exclplot}
\end{figure}

In Fig.~\ref{exclplot} we draw the 90\% C.L. contour defining
the sensitivity limit on $\theta_{13}$ in case of absence of a signal,
with $\delta_{CP}$ as a fixed free parameter. The sensitivity has been
computed applying a $\chi^2$ analysis including the expected
background and a 2\% systematic error. The sensitivity is rather good,
as can be argued from Fig.~\ref{exclplot}, but it is systematically
worse than the one of the baseline scenario.

\section{Conclusion}
The next generation of accelerator based neutrino oscillation
experiments has the challenging purpose to discover the missing
oscillation parameter $\theta_{13}$. The relevant role played by this
parameter in the neutrino oscillation physics and the wide
experimental program developed to discover it are related to its
strong correlation with the CP violation in the leptonic sector.
Indeed, a vanishing or too small value for $\theta_{13}$ would make
impossible the observation of the CP violation parameter $\delta$ and
of to fix the neutrino mass hierarchy.

In this paper we discussed in particular a neutrino program based on
the machine upgrades of the LHC. Indeed, it turns out that the
Super-SPS option for the luminosity/energy upgrade of the LHC has the
ideal features for the construction of a $\beta$-beam facility with a
$\gamma$ in the range $350-580$ whose physics case would be enormously
strengthened in the case of $\theta_{13}$ discovery in Phase I
experiments. Given that the luminosity/energy upgrade of the LHC is
foreseen after 2015, and that Phase I experiments are expected to
complete their program around that date, we see a window of
opportunity for a Phase II neutrino program in Europe compatible with
the LHC (and its upgrade) running.  This would allow, contrarily to
other proposed neutrino physics program, the full exploitation of
european accelerator facilities during the LHC era. Other advantages
are that the proposed experimental program does not imply the
construction of a Megaton detector, but of a very dense detector
(iron slabs interleaved with e.g. glass RPC segmented into
$2{\times}2~\mbox{cm}^2$ cells) with a few tens of kiloton mass. This
would fit into the underground facilities existing at the Gran Sasso,
whose distance from CERN, given the neutrino energy of this facility,
happens to be at the peak of oscillation probability!

The proposed detector will be able not only to identify $\nu_\mu$ and
$\bar{\nu}_\mu$ charged/current interactions, but also to measure the
energy of the incident neutrino. This opens, given the long baseline,
the possibility to measure the neutrino mass hierarchy. In case Phase
I experiments will discover a non vanishing $\theta_{13}$
($>3^\circ$), the proposed set-up will be able to discover CP
violation for $\delta$ values down to $30^\circ$. These performances
are comparable with the one obtained by the baseline $\beta$-beam that
foresee the construction of an accelerator complex with no overlap
with the LHC program and the excavation of a very large cavern able to
host a megaton water Cerenkov detector.  Note, however, that the
sensitivities at small values of $\theta_{13}$ of the medium $\gamma$
facilities result to be worse than the ones for the baseline
$\beta$-beam scenario, mainly due to the large difference of mass
(40 vs 1000 kton) of the corresponding detectors.

Finally, we want to point out that at the present, although very
promising, a detailed study of a $\beta$-beam complex is still
missing. There is an EURISOL design study group that is scrutinizing
the baseline option, but the medium $\gamma$ scenario is beyond its
scope. Nevertheless, the latter option surely deserves careful
consideration.


\end{document}